\documentclass[aps,prb,showpacs,twocolumn,floats,superscriptaddress,10pt]{revtex4-1}
\usepackage{graphicx}
\usepackage{dcolumn}
\usepackage{bm}
\usepackage{color}
\usepackage{amssymb}

\begin{document}
\title{Protection of a non-Fermi liquid by spin-orbit interaction}

\date{\today}

\author{T. K. T. Nguyen}  
\affiliation
{Institute of Physics, Vietnam Academy of Science and Technology,
10 Dao Tan, Hanoi, Vietnam}

\author{M. N. Kiselev}
\affiliation
{The Abdus Salam International Centre for Theoretical Physics, Strada
Costiera 11, I-34151, Trieste, Italy}

\begin{abstract}
We show that a thermoelectric transport through 
a Quantum Dot (QD) - single-mode Quantum Point Contact (QPC) nano-device 
demonstrating pronounced fingerprints of Non-Fermi Liquid (NFL) behaviour in the absence of external magnetic field is protected from magnetic field NFL destruction by strong spin-orbit interaction (SOI).
The mechanism of protection is associated with appearance of additional scattering 
processes due to lack of spin conservation in the presence of both SOI and small Zeeman field. 
The interplay between in-plane magnetic field $\vec B$ and SOI is controlled by the angle between 
$\vec B$ and $\vec B_{SOI}$.
We predict strong dependence of the thermoelectric coefficients on the orientation of the 
magnetic field and discuss a window of parameters for experimental observation of NFL effects.
\end{abstract}


\maketitle
\section{Introduction}
The paradigm of Landau Fermi Liquid (FL) \cite{LandauFL} is one of the cornerstones of modern condensed matter theory. Based on the concepts of quasiparticles -  well defined excitations whose energy in the long-wave limit is greater than their decay rate, the FL theory successfully explains the behaviour of normal and superconducting metals giving universal predictions for thermodynamic and transport properties \cite{Anderson}.  The FL phenomenology
is justified in many microscopic models describing  interacting fermions in- and out- of equilibrium. However, there are several
cases where a violation of the FL picture is observed experimentally (e.g. in strongly correlated electron systems such as heavy fermion compounds \cite{HF}, unconventional superconductors \cite{Anderson} and quantum transport through nano-structures \cite{Ralph, Goldhaber}). The pronounced Non-Fermi Liquid (NFL) behaviour of these systems is attributed to a breakdown the quasiparticle concept: the decay rate of low-energy excitations becomes greater than the energy of the excitations itself.  
\begin{center}
\begin{figure}[b]
 \includegraphics[width=1\columnwidth]{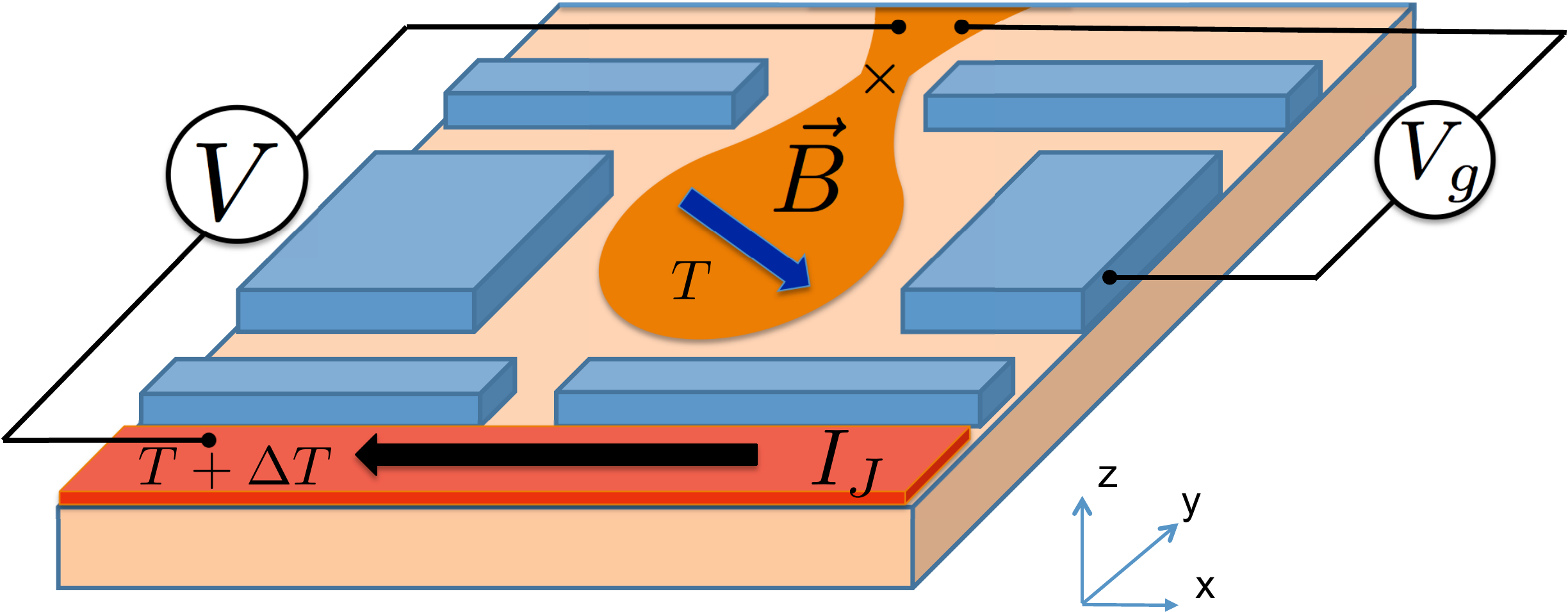} 
\caption{(Color online) Typical setup for thermoelectric measurements: "cold" contact (light orange area) at reference temperature $T$ - Quantum Dot - Quantum Point Contact electrostatically defined by gates (blue boxes), separated by a tunnel barrier from a "hot" (deep orange) contact at temperature $T+\Delta T$. The voltage $V$ is applied across the device for zero current measurements (see text for the details).}
\label{f.1} 
\end{figure}
\end{center}
\vspace*{-9mm}
While the fingerprints of NFL physics in thermodynamics
of strongly correlated systems and quantum transport had been seen experimentally \cite{HF, Goldhaber}, it is also generally accepted that the NFL picture is extremely sensitive to variation of external parameters being unstable against the FL ground state. Thus, the stability of the NFL domain and the possibility to observe strong deviations from the Landau FL paradigm posses major challenges including the development
of theoretical models and predictions for the stabilization of NFL-states. Important questions are: is it
possible at all to protect unstable NFLs? What are the physical observables which demonstrate the most pronounced manifestation of the NFL physics? 

In this paper we present an example based on a window of parameters within which the observation of
strong deviation from the FL picture can be protected and extended by effects of strong spin-orbit interaction (SOI). The physical observables we consider in this work are thermoelectric coefficients
of a nano-device (Fig. 1). Our theoretical model justifying NFL behaviour is a two-channel Kondo (2CK) model \cite{NozBla80,Zawad,Affleck91}. While the scattering of single orbital channel electrons on a resonance quantum impurity itself leads to strong modification of the thermoelectric transport properties within the Landau FL paradigm through strong renormalization of the FL energy scale \cite{Fazio,Costi2010,Zitko}, the detour from the FL picture is predicted to change completely both electric \cite{Matveev95,Flensberg,FM_theory,LeHur} and thermoelectric transports
\cite{MA_theory, NKK10}. For example, one of the manifestations of the NFL behaviour in quantum transport is associated with the logarithmic enhancement of the thermoelectric power \cite{MA_theory} in the situation when the 2CK model originates from the charge Kondo effect in a single mode quantum point contact (QPC) - quantum dot (QD) setup tuned by gate voltages to the Coulomb blockade (CB) peak regime \cite{Matveev95,Flensberg,FM_theory,LeHur,MA_theory}. In that case two channels are the electron spin degrees of freedom while the almost transparent QPC (weak back-scatterer) works as a quantum impurity. The 2CK physics is known to be unstable with respect to any effects which can potentially break (statically or dynamically) the symmetry between the channels \cite{Goldhaber,NKK10}. In particular, it has been shown that the effects associated with time-reversal symmetry breaking (TRS) due to an external magnetic field restore the FL properties at temperatures below $T_{\rm eff}$ tunable by the field \cite{NKK10}.
{\color{black} The universality class of unstable 2CK model than changes to single channel Kondo problem (1CK). {\it Fully screened} 1CK is characterized by stable {\it local } FL properties.} Therefore, while being very attractive from theoretical point of view, the 2CK physics suffers from serious experimental obstacles \cite{Ralph,Goldhaber,GG11} impeding a direct observation of NFL behaviour.

{\color{black}
The paper is organized as follows: we describe possible experimental setup for observing the NFL transport in Section II. A theoretical model accounting for the interplay between SOI,
external magnetic field and effects of Coulomb blockade in Quantum Dot is presented in Section III.
We discuss the solution of one-dimensional quantum-mechanical scattering problem in the presence of strong SOI in Section IV. An effective model describing a low-energy physics of the problem and its exact solution is presented in Section V. The transport coefficients computed with the help of the exact solutions are discussed in Section VI. Section VII is devoted to discussion of the key results of the paper including estimation for the parameters and definition of the conditions necessary for experimental observations of the NFL physics. Summary and Conclusions are given in the Section VIII. Details of derivation of the effective model are presented in the Appendix A.}

\section{Proposed experimental setup} 
We consider a two-terminal nano-device (see Fig. 1) designed to be used for thermoelectric measurements  
\cite{Mol1,Mol2}. The QD - QPC contains 2-d electron gas (2DEG) confined in $z$-direction 
(light orange area on the Fig. 1). The open QPC connects it to the drain at the reference temperature $T$. 
We assume that the Rashba SOI \cite{SOIbook, Spintronics}  (caused by the gradient of the confining potential in $z$-direction)
leads to appreciable effects which we will discuss in this paper.
The source is separated from the QD by a tunnel barrier with low transparency $|t|$$\ll $$1$. The temperature of the source (deep orange) is adjusted by the Joule heat controlled by the current $I_J$ flowing along the lead (black arrow). The temperature difference $\Delta T$ across the tunnel barrier is assumed to be small compared to the reference temperature $T$ to guarantee the linear response operation regime for the device. The QD is electrostatically controlled by two plunger gates (blue rectangles) to adjust the size of the electron island. The device is operated in the steady state of zero source-drain current 
$I_{sd}$$=$$0$$=$$G\Delta$$V_{th}$$+$$G_{T}$$\Delta T$, controlled by applying a thermo-voltage 
$\Delta V_{th}$ between the source and the drain. The QPC (denoted by the cross in the light orange area) is tuned to the single mode regime characterized by a controllable small reflectivity $|r|$$\ll $$1$. 
{\color{black} Under this assumption and neglecting the resistance of  the "metallic" QD we assume that the voltage difference $\Delta V_{th}$ arises across the tunnel barrier between the source and QD}. The transport coefficients: electric conductance  $G$ and thermoelectric coefficient $G_{T}$ (measured independently) define the thermoelectric power (TP) $S$:
\vspace{-1mm}
\[
G_{T}=\frac{\partial I_{sd}}{\partial \Delta T},\;\;\; G=\frac{\partial I_{sd}}{\partial V},\;\; S=-\left.\frac{\Delta V_{th}}{\Delta T}\right|_{I_{sd}=0}=\frac{G_{T}}{G}.
\]
We assume that the magnetic field (blue arrow) is applied parallel to the plane of 2DEG to avoid orbital effects.

\section{Theoretical model} 
The theoretical description of setup (see Fig 1) is formulated in terms of the Hamiltonian:
\begin{eqnarray}
H=H_{s}+H_{d}+H_{tun}+H_{z}.
\label{ham1}
\end{eqnarray}
Here $H_s$ and $H_d$ are the Hamiltonians of the source {\color{black} ("hot" contact)} and the drain {\color{black} ("cold" contact)}, respectively. $H_{tun}$ describes tunneling between the source and the drain and $H_z$ accounts for the Zeeman effect in both contacts. We assume that the source can be described by a standard FL approach
\begin{equation}
H_s=\sum_{k,\sigma}\epsilon_{k\sigma} c^\dagger_{k\sigma} c_{k\sigma}
\label{hs}
\end{equation}
here $c^\dagger$ and $c$ are creation/annihilation operators of quasi-particles {\color{black} (we adopt a system of units $\hbar=k_B=1$)}. The drain {\color{black} $H_d=H_c+H_{QPC}$} includes the Coulomb blockaded QD described by charging Hamiltonian $H_c$ and QPC
represented by 
\begin{equation}
H_{QPC}=H_{0}+H_{SOI}+H_{BS}
\label{hqpc}
\end{equation} 
We assume that the charge 
{\color{black} $\hat Q=e(\hat{n}_{s}+\hat{n}_{d})$} in  the QD is weakly quantized (mesoscopic CB regime 
\cite{AG}) and controlled by the gate voltage $V_g$:
\begin{eqnarray}
H_{c}=E_{c}\left[\hat{n}_{s}+\hat{n}_{d}-N(V_{g})\right]^{2},
\label{hamc}
\end{eqnarray}
here  {\color{black} $\hat{n}_{s}$ and $\hat{n}_{d}$} are the operators of the number
of electrons that entered the dot through the source and the drain respectively, 
{\color{black} $E_c\sim e^2/L_{QD}$}
is the charging energy of QD with geometric size  {\color{black} $L_{QD}$}. Below we ignore effects associated with finite mean-level spacing in the dot. While charge is only weakly quantized in the mesoscopic CB regime, the spin remains a good quantum number in the absence of SOI. However, when the SOI is present, two spin sub-bands are split {\it horizontally in k-space} and while spin is no more conserved, the sub-band index characterizes quantized states instead.
The single mode QPC being a short quantum wire can be viewed as 1-d electron system in the presence of Rashba SOI  \cite{GJPB05,Sun07,Miranda,Cheng} $H_{SOI}=\alpha_{R}\left[\vec{k}\times\vec{n}_{z}\right]\cdot\vec{\sigma}$:
\begin{eqnarray}
H_{0}=-iv_{F}\sum_{\lambda\sigma}\lambda\int_{-\infty}^{\infty}dy\Psi_{\lambda,\sigma}^{\dagger}\left(y\right)\partial_{y}\Psi_{\lambda,\sigma}\left(y\right),\label{hamLL} 
\end{eqnarray}
\vspace*{-5mm}
\begin{eqnarray}
H_{SOI}=
 \alpha_R k_F\sum_{\lambda\sigma}\lambda \int_{-\infty}^{\infty}dy 
 \left[ \Psi_{\lambda,\uparrow}^{\dagger}\Psi_{\lambda,\downarrow}
 +\Psi_{\lambda,\downarrow}^{\dagger}\Psi_{\lambda,\uparrow}\right].
 \label{hamsoi}
\end{eqnarray}
We denote here by $\Psi_{\lambda,\sigma}$ the left ($\lambda$$=$$-$) and right ($\lambda$$=$$+$) movers with spin $\sigma=\uparrow,\downarrow$. The constant $\alpha_R$ characterizes Rashba SOI strength. The $k_F$ and $v_F$$=$$k_F$$/$$m^*$ correspond to the Fermi momentum and Fermi velocity (here $m^*$ is a fermion's mass). The 1-d electron transport through the QPC is along the $y$-axis (see Fig.1). 
The Rashba SOI $H_{SOI}$$=$$\alpha_R$$k_y$$\sigma_x$ is associated with the electric field gradient along the $z$-axis and can be characterized by the effective SOI field 
$g$$\mu_B$$ \vec B_{SOI}$$/$$2$$=$$\alpha_R$$k_F$$\vec e_x$  perpendicular to the direction of electron transport ($g$ is the Lande factor, $\mu_B$ is the Bohr magneton). Notice, that the SOI field alone does not lead to the TRS breaking. 

The backscattering (BS) Hamiltonian describes a scattering of electrons with momentum transfer $2k_F$ on a  non-magnetic quantum impurity located at the origin and characterized by a short-range potential $V(y)$:
\begin{eqnarray}
H_{BS}=\sum_{\lambda,\sigma} \int d y \Psi_{\lambda,\sigma}^{\dagger}\left(y\right)V(y)\Psi_{\bar\lambda,\sigma}\left(y\right)e^{-2i\lambda k_F y}.
\label{hambc}
\end{eqnarray}
The Hamiltonian $H_{tun}$ represents the weak tunneling $|t_{k}|$$=$$|t|$$\ll $$1$ of the electrons from the left contact to QD:
\begin{eqnarray}
H_{tun}=\sum_{k\lambda\sigma}\left[t_{k} c^\dagger_{k\sigma} \Psi_{\lambda\sigma}(-\infty) + h.c.\right].
\label{hamtun}
\end{eqnarray}
The Zeeman Hamiltonian $H_z$  describes the effects of the external magnetic field  $H_z=-g\mu_B\vec B (\vec s_s+\vec s_d)$, where $\vec s_s$ and $\vec s_d$ are the spin densities of electrons in the source and drain respectively. We consider a situation when both sizes of the QD {\color{black} ($L_{QD}$)} and QPC
{\color{black} ($L_{QPC}$)} are small compared to the SOI length scale $L_{QD}\sim L_{QPC}$$\ll$$l_{SOI}$$=$$1$$/$$(m^*$$\alpha_R)$.
{\color{black} Since the effective energy scale determining the behaviour of the transport coefficients 
of the model (\ref{hs}-\ref{hamtun}) which will be referred below as the Kondo temperature $T_K$ is \cite{FM_theory} $\sim E_c$ (see Appendix A), the condition $l_{SOI}$$\gg $$L_{QD}$ is equivalent to $g$$\mu_B $$B_{SOI}$$\ll $$T_K$ (see a discussion about interplay between Kondo effect and SOI in Ref. \onlinecite{Kikoin12}).}
We also assume that the SOI effects in the QD are already taken into account by using the approach developed in Ref. \onlinecite{Aleiner01}.
\begin{center}
\begin{figure}[t]
 \includegraphics[width=1\columnwidth]{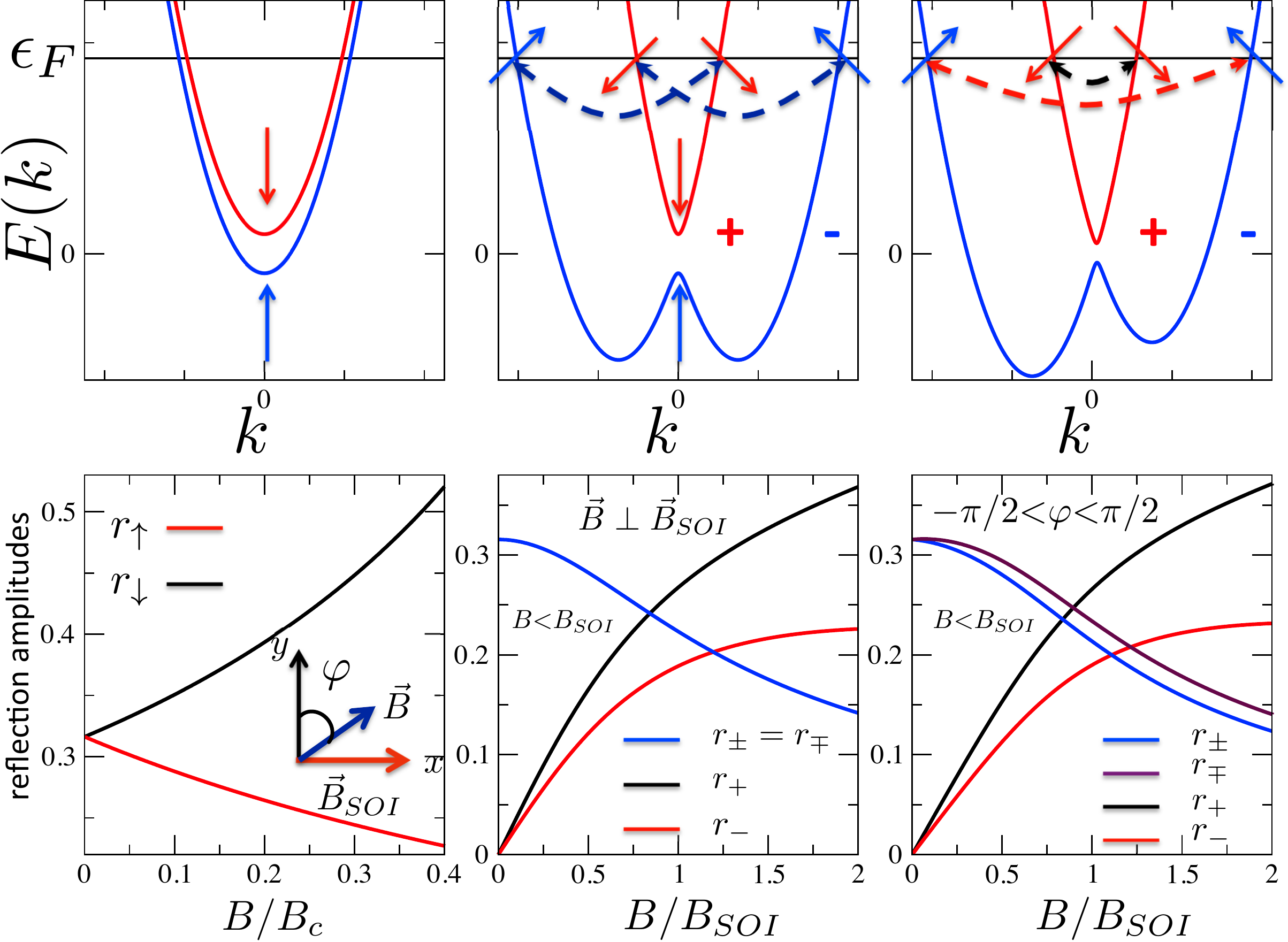} 
 \caption{(Color online) Top panels: two sub-band spectra (left) -  Zeeman splitting in the absence of SOI, (center) - $\vec B\perp\vec B_{SOI}$, angle between magnetic field and $y$-direction $\varphi$=$0$; (right)  - arbitrary angle  $-\pi/2$$<$$\varphi$$<$$\pi/2$. Bottom panels: magnetic field dependence of reflection amplitudes $|r_{\mu\nu}|$ for the spectra shown on top panels. {\color{black} For illustration we performed all calculations with model barrier $V(y)=V_0\exp(-|y|/L_{QPC})$,
$k_{0F} L_{QPC} = 3.6$ and hight of the barrier $V_0$ is tuned to get $r_0^2=0.1$ (see details in the text). Insert shows relative orientation of $\vec B$ and $\vec B_{SOI}$.}}
 \label{f.2} 
\end{figure}
\end{center}

\section{Scattering in the presence of SOI and magnetic field} 
{\color{black}
We consider 1d scattering problem in the presence of SOI \cite{SOIbook}
and Zeeman field applied parallel to the plane of 2DEG. The Hamiltonian is
given by 
\begin{equation}
{\cal H}={\cal H}_0 + V(y) = \frac{k^{2}}{2m^*}+\alpha_{R}\sigma_{x}k-\gamma\vec{\sigma}\cdot\vec{B}+V\left(y\right).
\label{hamm}
\end{equation}
The short-range potential $V(y)$ describes a non-magnetic impurity
located at the origin \cite{Miranda,Sun07}. The electron's transport
is along $y$-direction, $k$$=$$k_{y}$. Angle $\varphi$ characterizes
the orientation of magnetic field $\vec{B}$ with respect to $y$-
axis {\color{black} (see Fig. 2 left bottom panel insert)}.

The kinetic energy term of (\ref{hamm}) is given by 
\begin{eqnarray}
{\cal H}_{0}=\left(\begin{array}{cc}
\frac{k^{2}}{2m^*} & \alpha_{R}k+i\gamma Be^{i\varphi}\\
\alpha_{R}k-i\gamma Be^{-i\varphi} & \frac{k^{2}}{2m^*}
\end{array}\right).
\end{eqnarray}
The Hamiltonian ${\cal H}_{0}$ is promptly diagonalized in $k$-space.
The eigenvalues (spectra) describe two sub-bands ($\nu=+$ and $\nu=-$) split
both horizontally due to the SOI and vertically due to Zeeman effect (Fig. 2): 
\begin{equation}
E_{\pm}(k)=\frac{k^{2}}{2m^*}\pm\sqrt{\left(\alpha_{R}k\right)^{2}+\left(\gamma B\right)^{2}-2\alpha_{R}\gamma kB\sin\varphi}.\label{spectrum}
\end{equation}
(we use the short-hand notation $\gamma = g\mu_B/2$).
In the presence of both fields there are two sub-bands while the spin polarization changes continuously
as one moves from one Fermi point to the other along each sub-band. 
{\color{black} We assume that magnetic field is applied parallel to the plane of the 2DEG to decouple it from the orbital degrees of freedom and concentrate on the Zeeman effect only.} The angle $\varphi$ characterizes the orientation of $\vec B$ with respect to the axis of 1-d transport (y). The spectra
for $\varphi$$=$$0$ - perpendicular orientation of  $\vec B$ and $\vec B_{SOI}$ describe {\color{black} 
the situation when magnetic field $\vec B$ is oriented {\it along} direction of the transport.}
If $B$ is larger than $B_{SOI}$, the most important effects on thermoelectric transport are due to the Zeeman splitting of two sub-bands
(Fig 2 upper left panel) \cite{NKK10}. In that limit the effects of $B$ are associated with breaking of the channel symmetry, $|r_\uparrow|$$\neq $$|r_\downarrow|$ (Fig 2, lower left panel) which is crucial for the fate of NFL \cite{NKK10}.
For the case $B < B_{SOI}$ we distinguish two cases: i) $\varphi=0$ (Fig. 2, central panel) and ii) $-\pi/2$$<$$\varphi$$<$$\pi/2$, $\varphi\neq0$
(Fig. 2, right panel). {\color{black} Since the orbital effects are negligible for small magnetic fields
if $B < B_{SOI}$, the case i) $\varphi=0$ (Fig. 2, central panel) can also be realized when magnetic field is perpendicular to the plane of 2DEG. Besides, the theory discussed in the paper is also applicable when both Rashba and Dresselhaus SOI \cite{SOIbook, Spintronics} are present.  The transport coefficients for generic situation of in-plane $B$-field are fully determined by the angle $\Phi$$=$$\varphi_0$$-$$\varphi$ between $\vec B$ and $\vec B_{SOI}$, where $\varphi_0$ ($\varphi$) are the angles between $\vec{B}_{SOI}$ ($\vec B$) and axis of 1d motion respectively.} 
 
The eigenfunctions of $H_{1d}$ are momentum-dependent spinors $\Psi_{\nu}\left(y\right)=e^{ik\cdot y}\chi_{\nu}\left(k\right)$
\begin{eqnarray}
\chi_{\pm}\left(k\right)=\frac{1}{\sqrt{2}}\left(
\begin{array}{c}
\pm ie^{-i\vartheta\left(k\right)}\\
1
\end{array}\right),
\label{eigen}
\end{eqnarray}
where 
\begin{eqnarray}
\vartheta\left(k\right)=\arctan\left(\frac{\alpha_{R}k-\gamma B\sin\varphi}{\gamma B\cos\varphi}\right).
\end{eqnarray}
The four reflection amplitudes in the first order of the backscattering
potential are determined by $2k_{F}$ momentum transfer and given
by the matrix elements of $V(y)$ in spinor basis $\Psi_{\nu}$. The
diagonal matrix elements 
\begin{equation}
|r_{\mu\mu}|=\left|\frac{V\left(k_{F+}^{\mu}-k_{F-}^{\mu}\right)}{v_{0F}}\cos\left(\frac{\vartheta\left(k_{F+}^{\mu}\right)-\vartheta\left(k_{F-}^{\mu}\right)}{2}\right)\right|\label{rmm1}
\end{equation}
characterize the intra-band scattering (we shall use the short-hand notations
$|r_{\mu\mu}|$$\equiv $$|r_{\mu}|$ below). Here $k_{F+}^{\mu}$$>$$0$ and $k_{F-}^{\mu}$$<$$0$
stand for the right and left Fermi points of a sub-band $\mu$ respectively,
$v_{0F}$$\sim $$(m^*a)^{-1}$ originates from the high energy cutoff, $a$ is
a lattice constant. The off-diagonal matrix elements 
($|r_{+-}|$$\equiv $$|r_\pm|$ and $|r_{-+}|$$\equiv $$|r_\mp|$):
\begin{equation}
|r_{\mu\nu}|=\left|\frac{V\left(k_{F+}^{\mu}-k_{F-}^{\nu}\right)}{v_{0F}}\sin\left(\frac{\vartheta\left(k_{F+}^{\mu}\right)-\vartheta\left(k_{F-}^{\nu}\right)}{2}\right)\right|\label{rpm1}
\end{equation}
describe the inter-band scattering. 

{\color{black}
The backscattering Hamiltonian (\ref{hambc}) in the basis of eigenfunctions (\ref{eigen}) casts the following form:
\begin{eqnarray}
H_{BS}=v_{F}\sum_{\lambda\mu\nu}|r_{\mu\nu}|\left[\Psi_{\lambda,\mu}^{\dagger}(0)\Psi_{\bar{\lambda},\nu}(0)+h.c.\right].\label{FBSc}
\end{eqnarray}
Let us analyse various limits of the back-scattering
corresponding to different orientation of the in-plane magnetic field $\vec B$ in the regime of strong interplay with the effects of SOI.}

For the most generic case of interplay between SOI and Zeeman magnetic field there exist four independent scattering processes resulting in four different reflection amplitudes (Fig. 2, right lower panel).
The reflection amplitudes for the intra-band scattering 
$|r_+|$ and $|r_{-}|$ (black and red dashed arrow on Fig. 2, right upper panel) in the first order of the backscattering potential are proportional to the amplitude of $B$ (Ref. \onlinecite{Miranda})
\begin{eqnarray}
|r_{+/-}|=r_0 \left[\frac{k_{0F}}{k_{0F} \mp \delta}\right]\left(\frac{B}{B_{SOI}}\right)\cos\varphi,
\label{rpp}
\end{eqnarray}
where $k_{0F}$ is the Fermi momentum at zero splitting ($\delta$$=$$0$), $\delta$$=$$m^*\alpha_R$, $r_0\propto |V(2k_{0F})m^* a|\ll 1$ is a coefficient characterizing the transparency of the barrier.  
The intra-band scattering is completely suppresses for $\varphi$$=$$\pi$$/$$2$
since the angle $\vartheta(k_{F\pm})$$=$$\pm\pi/2$ and the eigenfunctions
do not depend on $B$. The intra-band reflection amplitudes $|r_{+}|$$=$$|r_{-}|$$=$$0$ while 
inter-band $(|r_\pm|$$,$$|r_\mp|)$$\neq $$0$. Thus, for $\varphi$$=$$\pi$$/$$2$
we have only two non-zero reflection amplitudes 
$|r_\pm|$ and $|r_\mp|$ and therefore the thermoelectric transport can be described by equations of Ref. \onlinecite{NKK10} if replacing 
$|r_\uparrow|$$\to $$|r_\pm|$ and $|r_\downarrow|$$\to $$|r_\mp|$.

The inter-band scattering amplitudes (blue dashed arrows on central upper panel of Fig. 2)  for the case $-\pi/2$$<$$\varphi$$<$$\pi/2$ are given by
\begin{eqnarray}
|r_{\pm/\mp}|=r_0\left[1\mp b\left(\frac{B\sin\varphi}{B_{SOI}}\right)-c(\varphi)\left(\frac{B}{B_{SOI}}\right)^2\right].
\label{rpm}
\end{eqnarray}
Here coefficients $(b,c(\varphi)) \sim 1$ depend on the geometry of the QPC. One can see that for $\varphi=0$, the scattering term linear
in $B$ (linear Zeeman effect) disappears and additional symmetry
$|r_\pm|=|r_\mp|$ emerges (Fig. 2 central panel). 
The reflection amplitudes depend on magnetic field quadratically
(quadratic Zeeman effect). Thus, the scattering Hamiltonian in that case contains  three independent scattering parameters.
}
\section{Effective model}
We recapitulate briefly the main steps of the derivation of transport coefficients (for details see Appendix A):
i) we bosonize the 1-d Hamiltonian (\ref{hamLL}-\ref{hambc}) using a standard approach \cite{Thierry, GNT}. The effective bosonic Hamiltonian gives us a boundary sine-Gordon (BSG) model \cite{GNT} with four different backscattering amplitudes. The high-$T$ results are obtained by perturbative expansion (in reflection amplitudes) around the strong-coupling fixed point of the model.
ii) The non-perturbative results in the low-$T$ regime are obtained by re-fermionization procedure through the mapping the BSG model onto the effective Anderson model \cite{Matveev95,MA_theory,LeHur,NKK10}. 
iii) The effects of the Zeeman field at the QPC resulting in
TRS breaking caused by the asymmetry of reflection amplitudes \cite{NKK10} are accounted by a magnetic field dependent {\color{black} resonance width $\Gamma$} at the CB peaks \cite{NKK10}. {\color{black} The
resonance width $\Gamma$ in the presence of Zeeman field remains finite for whole range of the gate 
voltages,} cuts the temperature-dependent logarithm and therefore restores FL properties. The width $\Gamma$ (Refs. \onlinecite{Matveev95,FM_theory,GNT}) is attributed to a {\it single local Majorana mode}
{\color{black} interacting with a {\it single} mode of chiral fermions} \cite{Matveev95,MA_theory} in the theory containing only two (intra-band) scattering processes.  The
{\color{black} scattering on} quantum impurity in the presence of SOI and Zeeman fields {\color{black} involves four "$2k_F$" processes} which can be accounted for by {\it two local Majorana modes} {\color{black} interacting with {\it four} modes of {\it two} species of the chiral fermions}. As a result, two different resonance widths enter the transport coefficients. The interplay between two widths associated with inter- and intra-band processes leads to remarkable effects in thermoelectric transport.

The effective Anderson model which describes a hybridization of two local Majorana fermions $\eta_1$ and $\eta_2$ with two species of {\color{black} {\it chiral}  fermions (see Appendix A) is a direct generalization of \cite{Matveev95, FM_theory, MA_theory, NKK10} for a case of interplay between Zeeman and SOI fields:}
\begin{eqnarray}
 &  & H_{\tau}(t)=\int_{-\infty}^{\infty}dk
 \left[\sum_{\alpha=1,2} (k\cdot v_{F}) c_{\alpha,k}^{\dagger}c_{\alpha,k}\right.\label{Kondo}\\
 &  & -\sqrt{2}\left.\left(\omega_{s\tau}(t)\eta_{1}(c_{1,k}-c_{1,k}^{\dagger})-i\omega_{a\tau}(t)\eta_{1}(c_{1,k}+c_{1,k}^{\dagger})\right.\right.\nonumber \\
 &  & \left.\left.+\omega_{ms\tau}(t)\eta_{2}(c_{2,k}-c_{2,k}^{\dagger})-i\omega_{ma\tau}(t)\eta_{2}(c_{2,k}+c_{2,k}^{\dagger})\right)\right].\nonumber 
\end{eqnarray}
{\color{black} where following Ref. \onlinecite{MA_theory} we define  
\begin{eqnarray}
\omega_{s/a\;\tau}(t) &=&\Omega_{s/a}\cdot f_{s/a\;\tau}(t),\nonumber\\ 
\omega_{ms/ma\;\tau}(t)&=&\Omega_{ms/ma}\cdot f_{s/a\;\tau}(t),
\label{om}
\end{eqnarray} 
with
\begin{eqnarray}
\Omega_{s/a} &=&\sqrt{\frac{v_F E_c e^C}{2\pi^3}}||r_+|\pm|r_-||,\nonumber\\
\Omega_{ms/ma}&=&\sqrt{\frac{v_F E_c e^C}{2\pi^3}}||r_\pm|\pm|r_\mp||,
\label{Omegas}
\end{eqnarray}
and the time-dependent functions
\begin{equation}
f_{s/a\;\tau} = (-1)^{n_\tau(t)} {\rm Re}/{\rm Im}\left[ \exp\left\{ i\left(\delta\chi_\tau(t) -\pi N\right)\right\}\right].
\end{equation}
Function $\delta \chi_\tau(t)$ describes the deviation of the phase of the charge mode mean value from $\pi n_\tau(t)$ (Ref. \onlinecite{MA_theory}):
\begin{eqnarray}
\delta \chi_\tau(t)\approx\frac{\pi^2 T}{2E_c}\left(\cot[\pi T (t-\tau)] -\cot[\pi T t]\right),
\label{chi}
\end{eqnarray}
here $N$ is a function of a gate voltage $V_g$: $N$ is integer in the Coulomb blockade valleys and half-integer in the Coulomb blockade peaks, $n_\tau(t)=\theta(t)\theta(\tau-t)$ ($\theta(t)$ is a step function) and  $C\approx 0.577$ is the Euler's constant.
}

{\color{black} The original Matveev model \cite{Matveev95} corresponds to the case $\Omega_{a}$$=$$\Omega_{ma}$$=$$\Omega_{ms}$$=$$0$ 
and describes a {\it single Majorana fermion} $\eta_1$ coupled to
the {\it odd} combination of creation/annihilation operators of {\it chiral} fermions. Contrast to conventional Anderson model which preserves $U(1)$ symmetry, the model \cite{Matveev95,FM_theory} is characterized by $Z_2$ symmetry instead. As a result, the NFL properties associated with the two-channel Kondo physics emerge. The NFL behaviour of the two-channel Kondo model are attributed to {\it overscreened} regime realized when the number of orbital channels ${\cal N}$ exceeds twice a spin of a quantum impurity. As it was shown in Ref. \onlinecite{NKK10}, Zeeman in-plane magnetic field restores the $U(1)$ symmetry through
appearance of non-zero $\Omega_a$ and therefore leads to the restoration of the  FL behaviour characteristic for the single-channel fully screened Kondo model in both thermodynamic  \cite{LeHur} and transport \cite{NKK10} coefficients. The Kondo temperature $T_K$ is of the order of the charging energy $E_c$ (see Appendix A). The effective Hamiltonian (\ref{Kondo}) describing scattering in the presence of both Zeeman and SOI fields has a structure of two copies of the two-channel Kondo model where coupling constants $\omega_{i\tau}$ depend on the magnetic field. Thus, when all reflection amplitudes are different, the model (\ref{Kondo}) is characterized by generic FL properties. However, if {\it accidental} degeneracy fine-tuned by the orientation of in-plane magnetic field appears, one of the non-identical copies of the two-channel Kondo model preserves the NFL properties.} 

The effects of interplay between the backscattering at the QPC and Coulomb interaction in the QD can be accounted by the correlator $K(\tau)=\langle T_\tau F(\tau)F^\dagger(0)\rangle$ (see details in 
Refs. \onlinecite{MA_theory, NKK10}). {\color{black} The operator $F(\tau)$ accounts for the {\it weak charge quantization} 
in the {\it mesoscopically Coulomb blockaded} QD. Following Ref. \onlinecite{MA_theory} we define the charge of QD $\hat Q=e(\hat n_\tau+\hat n_d)$, where $\hat n_\tau$ is an integer valued operator which commutes with the annihilation operator of the electron in the dot at the position of the source Ref. \onlinecite{FM_theory}.
Since by definition $[F(\tau),\hat n_\tau]=F(\tau)$, the role of operator $F(\tau)$ is to account for the effects of interaction in QD: $\Psi_\lambda(\tau)$$=$$F(\tau)\Psi_{0\lambda}(\tau)$ where $\Psi_\lambda$ and $\Psi_{0\lambda}$ correspond to interacting and non-interacting left/right fermions, respectively. Thus, the dressed Green's Function (GF) ${\cal G}(\tau)=-\langle T_\tau \Psi_{\lambda}(\tau)\Psi^\dagger_{\lambda}(0)\rangle$ and free fermionic Green's function
$G_0(\tau)$$=$$-\langle T_\tau \Psi_{0\lambda}(\tau)\Psi^\dagger_{0\lambda}(0)\rangle$$=$$-\pi \nu_0 T/\sin(\pi T \tau)$  are connected \cite{MA_theory} by simple relation ${\cal G}(\tau)=K(\tau)G_0(\tau)$ (here $\nu_0$ is a density of states in QD). The transport coefficients of the model are determined by the Green's function ${\cal G}$ (see next Section).}

{\color{black} In order to compute the Green's function ${\cal G}$ 
(or, equivalently, compute the correlator $K$)
we define the operator $U_\tau$$=$$\left(-1\right)^{d^{\dagger}d}$ where $d$$=$$($$\eta_1$$+$$i$$\eta_2$$)$$/\sqrt{2}$
and $d^\dagger$$=$$($$\eta_1$$-$$i$$\eta_2$$)$$/\sqrt{2}$.
We keep notations of the Matveev and Andreev work Ref. \onlinecite{MA_theory} for
$U_\tau$$=$$2i$$\eta_{2}$$\eta_{1}$.
Notice, that the "spin" and charge are completely disentangled in the correlator $K(\tau)$$=$$K_c(\tau)K_s(\tau)$.

While $K_c$$($$\tau$$)$$=$$\pi^2$$T$$e^{-C}$$/$$($$2$$E_c$$|$$\sin$$($$\pi$$T$$\tau$$)$$|)$ (see Ref. \onlinecite{FM_theory,MA_theory} for details of calculations), the $K_s(\tau)$ in zero-th order in $H_\tau(t)$$-$$H_{\tau=0}(t)$ is defined by the correlator
\begin{eqnarray}
K^{(0)}_s(\tau)=\langle T_{t}U(\tau)U(0)\rangle_{0}.\label{K0}
\end{eqnarray}
Here $\langle...\rangle_{0}$ denotes an averaging with (\ref{Kondo}) taken at $\tau$$=$$0$. 
{\color{black} (Notice obvious correspondence $\eta_1$$\to$$\sigma_x$$/$$\sqrt{2}$, 
$\eta_2$$\to$$\sigma_y$$/$$\sqrt{2}$ and $2$$i$$\eta_2$$\eta_1$$\to$$\sigma_z$, where $\sigma_i$ are spin $s$$=$$1$$/2$ operators ($i$$=$$x$$,y$$,z$). Therefore,
$K^{0}_s(\tau)$$=$$\langle$$T_t\sigma_z$$(\tau)$$\sigma_z$$(0)$$\rangle$.)}

The first non-vanishing order in $H_\tau(t)$$-$$H_{\tau=0}(t)$
correction to the correlator $K_s(\tau)$ is given by: 
\begin{eqnarray}
K^{(1)}_s(\tau)=-\int_{0}^{1/T}dt\langle T_{t}H'_{\tau}(t)U(\tau)U(0)\rangle_{0},\label{K1}
\end{eqnarray}
where the Hamiltonian $H'_{\tau}(t)$ has the form: 
\begin{eqnarray}
H'_{\tau}(t)= -2i\delta\chi_{\tau}(t)\left(\Omega_{s}\sin(\pi N)\eta_{1}\varsigma_{1}
 -\Omega_{a}\cos(\pi N)\eta_{1}\zeta_{1}\right.\nonumber 
\end{eqnarray}
\begin{eqnarray} 
\left.+\Omega_{ms}\sin(\pi N)\eta_{2}\varsigma_{2}
-\Omega_{ma}\cos(\pi N)\eta_{2}\zeta_{2}\right).
\end{eqnarray}
Here we define four additional Majorana fermions  ($\alpha$$=$$1,2$) 
through a $k$-Fourier transform of the even/odd combinations of {\color{black} creation/annihilation operators} of two species of {\color{black} {\it chiral} fermions $c_{\alpha,k}$}
taken at the position of the quantum impurity $y$$=$$0$:
\begin{eqnarray}
\zeta_{\alpha} & =\displaystyle\int_{-\infty}^{\infty}d k \zeta_{\alpha k}=\frac{1}{\sqrt{2}}\int_{-\infty}^{\infty} d k \left(c_{\alpha,k}+c_{\alpha,k}^{\dagger}\right),\nonumber\\
\varsigma_{\alpha} & =\displaystyle\int_{-\infty}^{\infty} d k\varsigma_{\alpha k}= 
\frac{1}{i\sqrt{2}}\int_{-\infty}^{\infty} d k \left(c_{\alpha,k}-c_{\alpha,k}^{\dagger}\right).
\label{condel}
\end{eqnarray}
In order to compute the correlators (\ref{K0}, \ref{K1}) we apply the Wick's theorem to the product of even number of fermions and express the result in terms of the products of the single particle GF's. 
The imaginary time (Matsubara) GF's form a $6\times 6$ matrix with 21 independent components
(6 diagonal and 15 off-diagonal): 
\begin{eqnarray*}
G_{\mu\nu}^{\eta\eta}(\tau) & = & -\langle T_{\tau}\eta_{\mu}(\tau)\eta_{\nu}(0)\rangle,\; G_{\mu\nu}^{\zeta\zeta}(\tau)=-\langle T_{\tau}\zeta_{\mu}(\tau)\zeta_{\nu}(0)\rangle,\\
G_{\mu\nu}^{\varsigma\varsigma}(\tau) & = & -\langle T_{\tau}\varsigma_{\mu}(\tau)\varsigma_{\nu}(0)\rangle,\; G_{\mu\nu}^{\zeta\varsigma}(\tau)=-\langle T_{\tau}\zeta_{\mu}(\tau)\varsigma_{\nu}(0)\rangle,\\
G_{\mu\nu}^{\zeta\eta}(\tau) & = & -\langle T_{\tau}\zeta_{\mu}(\tau)\eta_{\nu}(0)\rangle,\; G_{\mu\nu}^{\varsigma\eta}(\tau)=-\langle T_{\tau}\varsigma_{\mu}(\tau)\eta_{\nu}(0)\rangle.
\end{eqnarray*}
The GF 's of quadratic Anderson-type Hamiltonian (\ref{Kondo}) can be found exactly (e.g. by solving equations of motion for the Majorana fermions). For computing
the correlators (\ref{K0}, \ref{K1}) we need only 6 GF', namely, two diagonal local Majorana's GF
(here $R$ denotes the retarded GF's):
\begin{eqnarray}
G_{11,R}^{\eta\eta}(\epsilon)=\frac{1}{\epsilon+i\Gamma_{B}},\; G_{22,R}^{\eta\eta}(\epsilon)=\frac{1}{\epsilon+i\Gamma_{A}}\label{Gdiag}
\end{eqnarray}
and four off-diagonal hybridized GF's: 
\begin{eqnarray}
G_{11,R}^{\zeta\eta}(\epsilon) & = & \frac{\Omega_{a}\sin(\pi N)2\pi/v_{F}}{\epsilon+i\Gamma_{B}},\nonumber \\
G_{11,R}^{\varsigma\eta}(\epsilon) & = & \frac{\Omega_{s}\cos(\pi N)2\pi/v_{F}}{\epsilon+i\Gamma_{B}},\nonumber \\
G_{22,R}^{\zeta\eta}(\epsilon) & = & \frac{\Omega_{ma}\sin(\pi N)2\pi/v_{F}}{\epsilon+i\Gamma_{A}},\nonumber \\
G_{22,R}^{\varsigma\eta}(\epsilon) & = & \frac{\Omega_{ms}\cos(\pi N)2\pi/v_{F}}{\epsilon+i\Gamma_{A}}.\label{Goffdiag}
\end{eqnarray}
Here we denote the resonance widths associated with symmetric and antisymmetric combination of reflection amplitudes as:
\begin{eqnarray}
\Gamma_{s/ms} & = & \Omega_{s/ms}^{2}\cos^{2}(\pi N)4\pi/v_{F},\nonumber \\
\Gamma_{a/ma} & = & \Omega_{a/ma}^{2}\sin^{2}(\pi N)4\pi/v_{F}.\label{Gams}
\end{eqnarray}
Two resonance Kondo widths entering the transport coefficients are given by:
\begin{eqnarray}
\Gamma_{A}=\Gamma_{ms}+\Gamma_{ma},\;\;\;\;\;\Gamma_{B}=\Gamma_{s}+\Gamma_{a}.\label{GAGB}
\end{eqnarray}
Notice that ten GF's, namely $G_{\mu\nu}^{\zeta\zeta}$, $G_{\mu\nu}^{\varsigma\varsigma}$ and $G_{\mu\nu}^{\zeta\varsigma}$ do not depend on the local Majorana fermions describing the quantum impurity. These GF renormalise the correlations between the conduction electrons, but do not enter the Eqs. (\ref{K0}, \ref{K1}). Another five GF's allowed by the symmetry of the Hamiltonian (\ref{Kondo}) do not appear in the equations (\ref{K0}, \ref{K1}) due to specific form of fermionic correlations in the Hamiltonian $H'_\tau(t)$.\\

\section{Transport coefficients}
The thermoelectric coefficient $G_T$ and electric conductance $G$ 
\begin{eqnarray}
G_T &=&-\frac{i\pi^2 G_L T}{2e}\int_{-\infty}^{\infty}\frac{\sinh(\pi T t)}{\cosh^3(\pi T t)}K\left(\frac{1}{2T}+it\right) dt,\;\;\;\;\;\;\;\\
G &=& \frac{ \pi G_L T}{2}\int_{-\infty}^{\infty}\frac{1}{\cosh^2(\pi T t)}K\left(\frac{1}{2T}+it\right) dt
\label{gg}
\end{eqnarray}
are here calculated by accounting for interaction effects in the QD through the correlator 
$K(\tau)$ {\color{black} defined in the previous Section}. The conductance of {\color{black} the barrier between the source and QD} $G_L=2\pi e^2 \nu_0\nu_L |t|^2$ is expressed through Fermi's golden rule as function of
the density of states (DoS) of the {\color{black} source} $\nu_L$, the DoS of the QD $\nu_0$ and the weak tunnelling amplitude $|t|$ (Ref. \onlinecite{comSOI}).

The correlator {\color{black} $K^{(0)}_s(1/(2T)+it)$} defined by (\ref{K0})
is an even function of time. This correlator determines the behaviour
of the electric conductance $G$, but does not contribute to $G_{T}$:
\begin{eqnarray}
G^{(0)}=\frac{G_{L}\Gamma_{A}\Gamma_{B}e^{-C}}{32\pi TE_{c}}F_{G}\left(\frac{\Gamma_{A}}{T},\frac{\Gamma_{B}}{T}\right).\label{G0}
\end{eqnarray}

\begin{widetext}
The equation for the thermoelectric coefficient $G_{T}$ is given
by an odd function {\color{black} $K^{(1)}_s(1/(2T)+it)$} defined by (\ref{K1}):
\begin{equation}
G_{T}^{\left(1\right)}=-\frac{G_{L}\Gamma_{A}\Gamma_{B}\sin\left(2\pi N\right)}{6e\pi^{2}E_{c}}\left[\frac{{\color{black}|r_+r_-|}}{\Gamma_{B}}\ln\left(\frac{E_{c}}{T+\Gamma_{B}}\right)F\left(\frac{\Gamma_{B}}{T},\frac{\Gamma_{A}}{T}\right)
+\frac{{\color{black}|r_\pm r_\mp|}}{\Gamma_{A}}\ln\left(\frac{E_{c}}{T+\Gamma_{A}}\right)F\left(\frac{\Gamma_{A}}{T},\frac{\Gamma_{B}}{T}\right)\right].\label{G1}
\end{equation}
The ratio of $G_{T}$ and $G$ defines the thermoelectric power: 
\begin{equation}
S=-\frac{16e^{C}\sin\left(2\pi N\right)T}{3e\pi}\left[\frac{{\color{black}|r_+r_-|}}{\Gamma_{B}}\ln\left(\frac{E_{c}}{T+\Gamma_{B}}\right)\frac{F\left(\frac{\Gamma_{B}}{T},\frac{\Gamma_{A}}{T}\right)}{F_{G}\left(\frac{\Gamma_{A}}{T},\frac{\Gamma_{B}}{T}\right)}
+\frac{{\color{black}|r_\pm r_\mp|}}{\Gamma_{A}}\ln\left(\frac{E_{c}}{T+\Gamma_{A}}\right)\frac{F\left(\frac{\Gamma_{A}}{T},\frac{\Gamma_{B}}{T}\right)}{F_{G}\left(\frac{\Gamma_{A}}{T},\frac{\Gamma_{B}}{T}\right)}\right].\label{TP}
\end{equation}
The functions $F_G$ and $F$ universally depend on the ratio of the resonance Kondo widths $\Gamma_A$, $\Gamma_B$ and the temperature:
\begin{eqnarray}
F_{G}\left(x,y\right)=\int_{-\infty}^{\infty}\int_{-\infty}^{\infty}dzdz^{\prime}\frac{\left[\left(z+z^{\prime}\right)^{2}+\pi^{2}\right]}{\left[\left(z^{\prime}\right)^{2}+x^{2}\right]\left[z^{2}+y^{2}\right]}\frac{1}{\cosh\left(\frac{z}{2}\right)\cosh\left(\frac{z^{\prime}}{2}\right)\cosh\left(\frac{z+z^{\prime}}{2}\right)},\label{FG}
\end{eqnarray}
\begin{eqnarray}
F\left(x,y\right)
=\int_{-\infty}^{\infty}\int_{-\infty}^{\infty}dzdz^{\prime}\frac{z\left(z+z^{\prime}\right)\left[\left(z+z^{\prime}\right)^{2}+\pi^{2}\right]}{\left[z^{2}+x^{2}\right]\left[\left(z^{\prime}\right)^{2}+y^{2}\right]}\frac{1}{\cosh\left(\frac{z}{2}\right)\cosh\left(\frac{z^{\prime}}{2}\right)\cosh\left(\frac{z+z^{\prime}}{2}\right)}.\label{F}
\end{eqnarray}
\end{widetext} 

\begin{center}
\begin{figure}[t]
 \includegraphics[width=1\columnwidth]{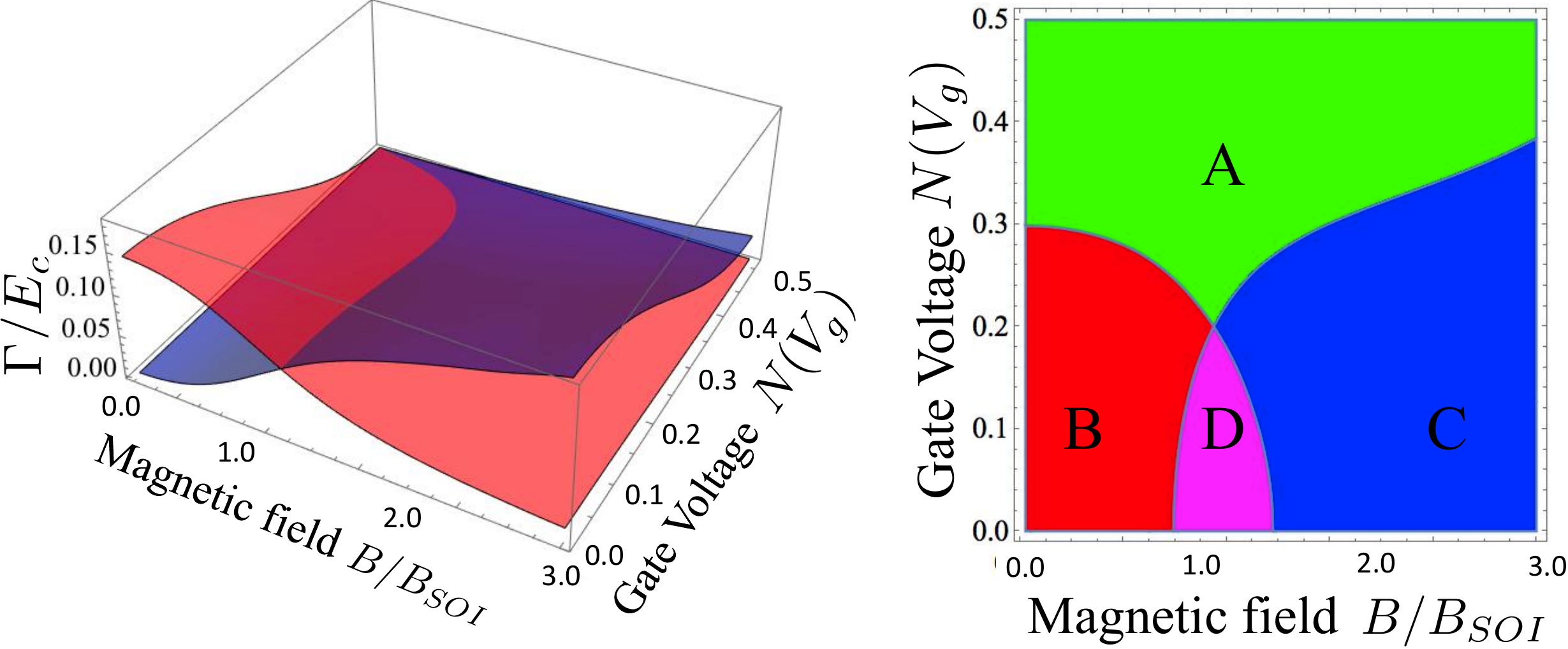} 
 \caption{{\color{black}(Color online) Left panel: Gate voltage and magnetic field dependence of $\Gamma_{\alpha}/E_c$: red - $\Gamma_A$, blue - $\Gamma_B$. Right panel: a "phase diagram" - four main regimes of the  thermoelectric transport are inside the green, red, blue and magenta domains; A - perturbative NFL, B - weak partial NFL, C- strong partial NFL, D - non-perturbative FL (see details in Section VII). Domains boundaries are defined by the {\it crossover} conditions $\Gamma_A(B,N)$$=$$T$ and $\Gamma_B(B,N)$$=$$T$.  For all plots $\alpha_R$$=$$0.15$$v_F$, $\varphi$$=$$\pi/4$, $r_0^2$$=$$0.1$, $k_{0F}L_{QPC}$$=$$3.6$, $T$$=$$0.05E_c$.}}
 \label{f.3n} 
\end{figure}
\end{center}
\vspace*{-18mm}
\section{Results and discussion} 
\subsection{Four main regimes of thermoelectric transport}
The non-perturbative equation for the resonance width $\Gamma$ related to the inter-band scattering (\ref{Gams}-\ref{GAGB}) demonstrates a weak dependence of $\Gamma$ on the magnetic field away from the CB peaks {\color{black}(see Fig.3 left panel)}:
\begin{eqnarray}
\Gamma_{A} \propto \Gamma_0\left[\left(1-\Lambda_A^2\right)\cos^2(\pi N) 
+ \Delta_A^2\right],
\label{GammaA}
\end{eqnarray}
where $\Gamma_0$$=$$r_0^2$$E_c$,  $\Delta_A(B,\varphi)$$=$$b$$\left(B/B_{SOI}\right)\sin\varphi$ and
$\Lambda_A^2=\Delta_A^2+2c(\varphi)(B/B_{SOI})^2$.

In contrast, the resonance width $\Gamma$ associated with the intra-band scattering (\ref{Gams}-\ref{GAGB}) 
strongly depends on $B$ at all gate voltages:
\begin{eqnarray}
\Gamma_{B} \propto \Gamma_0\left[(1 -  \Delta^2_B)\cos^2(\pi N)+ 
\Delta^2_B\right]\left(\frac{B\cos\varphi}{B_{SOI}}\right)^2.
\label{GammaB}
\end{eqnarray}
The {\color{black} $\Gamma^{min}_{B}$$\equiv$$\Gamma_B$$($$N$$=$$\frac{1}{2}$$)$$\propto$$\Gamma_0$$\cdot $$(B\cos\varphi/B_c)^2$} is a minimal resonance width, $\Delta_B$$=$$\delta/k_{0F}$ and $B_c$$\sim $$D$, where $D$$\sim $$(m^*a^2)^{-1}$ is the bandwidth. {\color{black} Thus, $B_c$ corresponds to the field
strength that is necessary to reach full  spin polarization of the conduction channel.}

Varying the temperature, gate voltage, amplitude and direction of the magnetic field one can achieve four different regimes of thermoelectric transport {\color{black} (Fig. 3 right panel)}:\\
A) $(\Gamma_{A},\Gamma_{B})$$\ll $$T$ - fully perturbative {\it NFL} regime. 
While $\Gamma_{B}$ is gapped and the gap is
$\Gamma^{min}_B$$\sim $$B^2$, $\Gamma_{A}$ could be gapless if the gate voltage is fine-tuned to the positions of CB peaks $N$$\to$$1$$/$$2$ and $\varphi$$\to $$0$. The TP (\ref{G0}-\ref{TP}) demonstrates fingerprints of {\it weak} NFL logarithmic behaviour:
\begin{eqnarray}
S\propto r_0^2\ln\left(\frac{E_c}{T}\right)\sin(2\pi N).
\label{SPT}
\end{eqnarray}
B) $\Gamma_{B}$$\ll$$T$$\ll$$\Gamma_{A}$ - perturbative in $\Gamma_{B}$$/T$ and non-perturbative
in $\Gamma_{A}$$/$$T$ (see (\ref{G0}-\ref{TP})). This regime can be reached either by fine-tuning the gate voltage away from the CB peaks or by tuning the direction of Zeeman field to be parallel to SOI in order to suppress the intra-band scattering:
\begin{equation}
S\propto \left[|r_{+}r_{-}|\ln\left(\frac{E_c}{T}\right)+
|r_{\pm}r_{\mp}| \frac{T}{\Gamma_{A}}
\ln\left(\frac{E_c}{\Gamma_{A}}\right)\right]\sin(2\pi N).
\end{equation}
The {\it weak} NFL effects are manifested in the TP log- behaviour originated from the intra-band scattering. The inter-band processes result in appreciable FL corrections to the TP (\ref{G0}-\ref{TP}). The NFL effects are {\it weak} 
since the amplitude of intra-band scattering is small at $B< B_{SOI}$.\\
C) $\Gamma_{A}$$\ll $$T$$\ll $$\Gamma_{B}$ - perturbative in $\Gamma_{A}$$/$$T$ and non-perturbative
in $\Gamma_{B}$$/$$T$ (see \ref{G0}-\ref{TP}). This regime is achieved in the vicinity of CB peaks and characterized by
{\it strong NFL} effects due to a weak magnetic field dependence of 
$|r_\pm|$ and $|r_\mp|$ protected by SOI. Thus, by fine-tuning the orientation of magnetic field perpendicular to SOI $\varphi$$=$$0$ one can controllably protect the NFL behaviour of TP in the regime $B$$<$$B_{SOI}$. 
The magnetic field controlled gap associated with the intra-band scattering weakly
depends on the gate voltage and results in small FL corrections to TP (compared to NFL effects):
\begin{equation}
S\propto \left[|r_{+}r_{-}|\frac{T}{\Gamma_{B}}\ln\left(\frac{E_c}{\Gamma_{B}}\right)+
|r_{\pm}r_{\mp}|
\ln\left(\frac{E_c}{T}\right)\right]\sin(2\pi N).
\end{equation}
D) $T$$\ll $$(\Gamma_{A},\Gamma_{B})$ - FL non-perturbative regime. The NFL logs associated with the intra- and inter- band scattering processes are cut by the corresponding resonance widths (\ref{G0}-\ref{TP}): 
{\color{black}
\begin{equation}
S\propto T\left[\frac{|r_{+}r_{-}|}{\Gamma_{B}}\ln\left(\frac{E_{C}}{\Gamma_{B}}\right)+
\frac{|r_{\pm}r_{\mp}|}{\Gamma_{A}}\ln\left(\frac{E_{C}}{\Gamma_{A}}\right)\right]\sin\left(2\pi N\right)
\label{TPFL}
\end{equation}
The TP is linear function of the temperature. However, the coefficient in front of $T$ strongly depends on
both gate voltage and magnetic field. 
}

\begin{center}
\begin{figure}[t]
 \includegraphics[width=1\columnwidth]{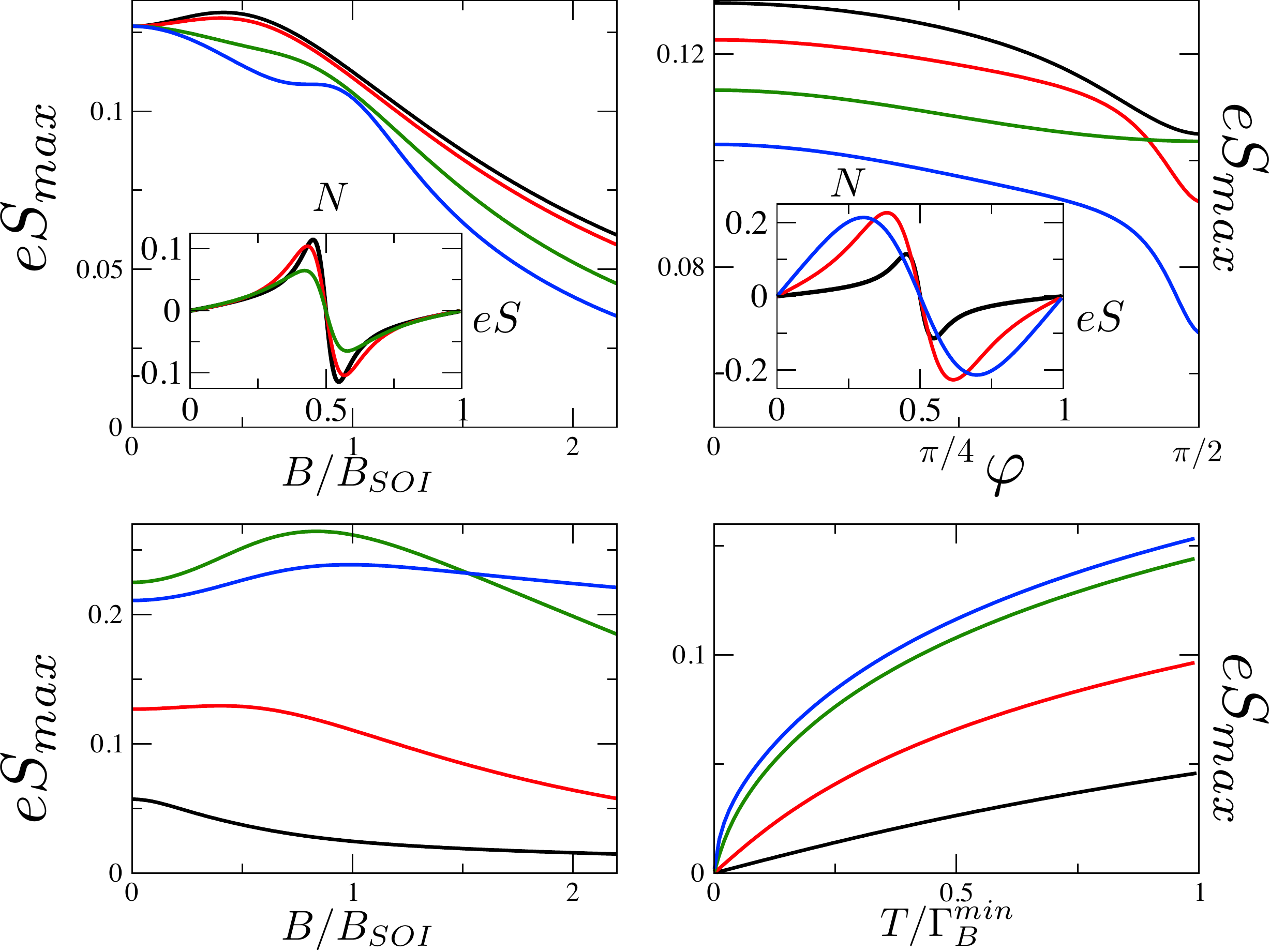} 
 \caption{{\color{black}(Color online) Top left: $eS_{max}$ as function of $B/B_{SOI}$ for different 
angles $\varphi$ at $T$$=$$0.001E_c$ and $\alpha_R$$=$$0.15v_F$, from top to bottom $\varphi$$=$$\pi/12$$,$$\pi/6$$,$$\pi/3$$,$$5\pi/12$. Insert: $eS(N)$ for $\varphi$$=$$5\pi/12$ for $B$$=$$0.5B_{SOI}$
(black), $B$$=$$B_{SOI}$ (red), $B$$=$$1.5B_{SOI}$ (green). Top right: $eS_{max}$ as function of the angle
$\varphi$ for different amplitudes of $B/B_{SOI}$ at $\alpha_R$$=$$0.15$$v_F$ and $T=0.001 E_c$ from top to bottom: $B/B_{SOI}=$$0.6,0.8,1.0,1.2.$. Insert: $eS(N)$ for $\varphi$$=$$5\pi/12$, $B$$=$$0.5B_{SOI}$,
$T$$=$$0.001$ (black), $0.01$ (red), $0.1$ (blue).} {\color{black} Bottom left:  $eS_{max}$ as function of $B/B_{SOI}$ for different  $T/E_c$ at $\alpha_R$$=$$0.15v_F$ and $\varphi$$=$$\pi/6$; 
$T/E_c$$=$$10^{-1}$- blue (regime $A$), $10^{-2}$- green (crossover $A$$\to$$C$), $10^{-3}$- red ($B$$\to$$D$$\to$$C$), $10^{-4}$ - black ($B$$\to$$D$) .} {\color{black} Bottom right: $e$$S_{max}$ as a function of $T/$$\Gamma_B^{min}$ ($\Gamma_B^{min}$$=$$\Gamma_B$$($$1/2)$) with $a_R$$=$$0.15$$v_F$, $B$$=$$0.5$$B_{SOI}$, for $\varphi$$=$$5\pi/12$ (black), $\varphi$$=$$\pi/3$ (red), $\varphi$$=$$\pi/6$ (green), $\varphi$$=$$\pi$$/12$ (blue), ($r^2_0$$=$$0.1$ and $k_{0F}$$L_{QPC}$$=$$3.6$).}}
 \label{f.3} 
\end{figure}
\end{center}
\vspace*{-8mm}
\subsection{Possible experimental realization and "smoking gun" predictions}
{\color{black}
{\it \underline{Choice of a material.}}  We suggest to use narrow-gap semiconductors, e.g. $InSb$ or $InAs$ for the observation 
of the NFL fingerprints in the quantum transport. Both materials are characterized by large bulk $g$-factors, e.g. $|g|\sim 10$ in InAs (see Ref. \onlinecite{Petta11})  and $|g|\sim 50$ in InSb (see Ref. \onlinecite{Mourik12}). The domain of parameters favourable for the observation of the NFL regime is defined as: $\delta_0 < T < \gamma B_{SOI} < E_c < \epsilon_F$,
where $\delta_0 \sim 1/(\nu_0 V_{QD})$ is a single-particle mean-level spacing in the QD of the size $L_{QD}$ and "volume" $V_{QD}$: $\delta_0^{2D}\sim \hbar^2/(m^{*} L_{QD}^2)$ and $\delta_0^{3D}\sim 
(k_F L_{QD})^{-1}\cdot\hbar^2/(m^{*} L_{QD}^2)$, $m^*$ is an effective mass of the carrier in a semiconductor, other parameters are defined in the previous Sections. The condition $L_{QD} < l_{SOI}$
allows to disregard the effects of the SOI in the QD (Ref. \onlinecite{Aleiner01}), while the condition
$L_{QPC} < l_{mfp}$
defines a ballistic regime of quantum transport through the QPC (here $l_{mfp}$ is an elastic
mean-free path). 

According to Ref. \onlinecite{Mourik12}, the parameters for $InSb$ QD - QPC
are as follows: $\hbar \alpha_R \sim 0.1 eV\cdot \AA \div 0.2 eV\cdot \AA$, 
$l_{SOI}=\hbar/(m^*\alpha_R)\sim 200 nm\div 400 nm$,
$|m^*| \sim 0.015 m_e$ ($m_e$ is electron's mass), $E_{SOI}=m^*\alpha_R^2/2 \sim 50 \mu eV$, the typical charging energy $E_c \sim 1 meV$
and typical Fermi velocities are $\hbar v_F \sim 0.5 eV \cdot \AA \div 1 eV \cdot \AA$, while the mean level spacing $\delta_0 < 10 \mu eV$ for $L_{QD} \sim l_{SOI}$. The mean-free path of the QPC of a width
$d_{QPC}\sim 10nm$ is $l_{mfp}\sim 300 nm \div 1\mu m$. Typical parameters for the $InAs$ 
QD - QPC are not much different \cite{Petta11}: $|m^*| \sim 0.03 m_e$, $\hbar \alpha_R \sim 0.05 eV\cdot \AA \div 0.3 eV\cdot \AA$, $l_{SOI} \sim 200 nm \div 1\mu m$ and $l_{mfp}\sim 300 nm \div 1\mu m$. Therefore, if we assume that $L_{QD}\approx L_{QPC} \sim 300 nm\div 500 nm$, our predictions could be verified at magnetic fields $B < 500 mT$ and temperatures $T\sim 100 mK \div 300 mK$ for typical densities of 2DEG $n_{2DEG}\sim 10^{11} cm^{-2}\div  10^{12} cm^{-2}$.
This estimation for parameters is taken from available literature (to our best knowledge), but may vary due to anisotropic character of the $g$-factor which in turn depends on external magnetic field \cite{Petta11} and may also be strongly reduced in confined geometries of the nano-structures.

{\it \underline{Testing a Mott-Cutler law.}} The first important test of the interplay between effects of the SOI and Zeeman field is to verify the Mott-Cutler (MC) law \cite{Cutler69} at external in-plane magnetic field. The MC-law is a standard benchmark for the FL properties \cite{MA_theory}. The MC-law says that the TP is proportional to a log-derivative of the electric conductance with respect to a position of the chemical potential (gate voltage $V_g$):
\begin{equation}
S \propto \frac{T}{E_c} \frac{\partial \ln G}{\partial N(V_g)}
\end{equation}
In the limit $T\ll(\Gamma_{A},\Gamma_{B})\ll E_c$ corresponding to the FL regime we get:
\begin{equation}
\frac{\partial\ln G}{\partial N(V_g)}\propto  E_c \left[\frac{|r_{+}r_{-}|}{\Gamma_{B}}+
\frac{|r_{\pm}r_{\mp}|}{\Gamma_{A}}\right]\sin\left(2\pi N\right)
\end{equation}
while the TP is given by (\ref{TPFL}).
Thus, a strong deviation from the MC-law in the FL regime at finite in-plane magnetic fields 
is a pre-cursor for the NFL behaviour
discussed in the paper. Notice that break down of the MC-law indicates that there exists no equivalent 
classic electric circuit consisting of the resistances connected in parallel or in series and therefore
the effects of both intra- and inter-band scattering play an important role in the quantum transport.
The violation of MC-law in the thermo-electric transport through a single-electron transistor has been reported in Ref. \onlinecite{Mol1}. We are not aware of existence of theoretical explanation of this effect in the framework of the FL theory.

{\it \underline{Thermopower in the presence of the external B-field.}} The next step is to measure 
the thermopower of a prototype nano-device (Fig.1). The magnitude and orientation of the in-plane magnetic field can be controlled in a standard way by four magnetic coils (not shown in the picture).
The TP maximum $eS_{max}(B)$ demonstrates a non-monotonic magnetic field dependence (strong NFL) which is most pronounced when $\vec B$ is orthogonal to $\vec B_{SOI}$ (black curve on Fig. 4 top left and right panels.) This has to be contrasted to almost monotonic TP maximum behaviour (blue curves) characteristic for weak NFL - FL regimes. The non-monotonic behaviour of TP maximum as a function of magnetic field is a central prediction of our paper. The non-monotonicity indicates that the NFL regime of TP is protected by SOI contrast to FL-like behaviour demonstrating rapid decrease of TP when magnetic field increases \cite{NKK10}.
Another indication of the NFL behaviour is attributed to the gate voltage dependence (Fig. 4 inserts). According to \cite{MA_theory, NKK10} it is characterized by strongly non-sinusoidal form (Fig. 4 inserts). The TP maximum at zero field, NFL regime, scales according to \cite{MA_theory} as 
$e S_{max}$$\sim $$r_0$$\sqrt{T/E_c}$$\ln(E_c/T)$. 
The TP maximum in the FL regime scales as $eS$$ \sim $$T/T_{\rm eff}$ with 
$T_{\rm eff}$$/E_c$$=$$B$$/$$B_c$$\ln^{-1}(B_c/(B|r_0|)$ (Ref. \onlinecite{NKK10}).} 
{\color{black} The $B$-field dependence of TP maximum measured at different temperatures 
(Fig. 4 left bottom panel) allows to distinguish between four main regimes A-D
discussed in the previous subsection. This measurement can be used for identification of crossovers between different domains. The TP maximum depends linearly on $T$ in the FL regime for $T$$<$$\Gamma^{min}_B$$=$$\Gamma_B$$($$N$$=$$1$$/2$$)$ (Ref. \onlinecite{NKK10}). This regime holds for $\varphi$$\to$$\pi$$/$$2$. In contrast to FL regime, the temperature dependence of TP maximum pronouncedly departs from the linear
behaviour (see Fig. 4 right bottom panel) when $\varphi$ is detuned from $\pi/2$. We suggest to test experimentally this effect as a benchmark for the NFL physics.}
\vspace*{-3mm}
\section{Summary and conclusions}  
\vspace*{-3mm}
We have demonstrated that the theory describing scattering of electrons characterized by two orbital degrees of freedom on a spin $s$$=$$1/2$  quantum impurity (two channel Kondo model) is strongly modified in the presence of both appreciable spin-orbit interaction and Zeeman splitting. It is shown that, on the one hand, the lack of spin conservation due to SOI leads to the appearance of new (extra) scattering channels which potentially enhance the thermoelectric transport. On the other hand, the Zeeman splitting
produces non-zero resonance widths of Majorana modes describing the quantum impurity and thus suppresses the NFL effects. The interplay between these two tendencies can be controlled by fine-tuning the angle between Zeeman and SOI fields. Our calculations predict a strong dependence of the thermoelectric power on the angle between $\vec B$ and $\vec B_{SOI}$ and thus open a possibility to  control the scattering mechanism by changing between four, three or two independent scattering processes. While the cases of four- and two- weak back-scattering do favour the FL behaviour, the additional degeneracy in scattering amplitudes appearing for three- scattering models due to SOI effects protects the NFL behaviour for the range of magnetic fields $B$$<$$B_{SOI}$. We conclude therefore, that SOI can indeed protect the NFL
against the destructive effects associated with breaking of channel symmetry.
\vspace*{-3mm}
\section{Acknowledgements} 
\vspace*{-3mm}
We are grateful to J.C. Egues, V. Fal'ko,  L. Glazman, K. Kikoin,  A. Komnik,  S. Ludwig, C. Marcus, K. Matveev,  L.W. Molenkamp and O. Starykh  for illuminating discussions. T.K.T.N acknowledges support through short-term visiting program of ICTP and Vietnam National Foundation for Science and Technology
Development (NAFOSTED) under grant number 103.01-2014.24. 
\vspace*{-3mm}
\begin{appendix}
\vspace*{-3mm}
\section{Effective Hamiltonian}
\subsection*{Backscattering: from fermions to bosons}
The backscattering Hamiltonian mixes the left- and right- moving fermions:
\begin{eqnarray}
H_{BS}=v_{F}\sum_{\lambda\mu\nu}|r_{\mu\nu}|\left[\Psi_{\lambda,\mu}^{\dagger}(0)\Psi_{\bar{\lambda},\nu}(0)+h.c.\right].\label{FBS}
\end{eqnarray}
The Hamiltonians Eqs. (3-4) of the main text and the Hamiltonian (\ref{FBS}) can be bosonized \cite{Thierry, GNT} in terms of dual fields
$\phi_{\nu}(y)$ and $\theta_{\nu}(y)$ satisfying commutation relations $\left[\phi_{\nu}(y),\theta_{\mu}(y')\right]=-i\pi\delta_{\nu\mu}\text{{sgn}}\left(y-y'\right)/2$ (Refs. \onlinecite{Thierry, GNT}):
\begin{eqnarray}
\Psi_{\lambda,\nu}(y)=\frac{u_{\lambda,\nu}}{\sqrt{2\pi a}}e^{i\lambda k_{F}^{\nu}y}\exp\{i[-\lambda\phi_{\nu}(y)+\theta_{\nu}(y)]\},
\end{eqnarray}
where $u_{\lambda,\nu}$ are Klein factors \cite{Thierry,GNT} introduced
to ensure proper anticommutation relations between the right- and left-
moving fermions.

{\color{black} Using a standard procedure \cite{Thierry,GNT}} we introduce the symmetric (charge) and antisymmetric ("spin") dual variables 
$\phi_{c,s}(y)$$=$$[\phi_{+}(y)$$\pm$$\phi_{-}(y)]$$/$$\sqrt{2}$
and $\theta_{c,s}(y)$$=$$[\theta_{+}(y)$$\pm$$\theta_{-}(y)]$$/$$\sqrt{2}$
satisfying the commutation relations 
$\left[\phi_{c/s}(y),\theta_{c/s}(y')\right]$$=$$-i$$\pi$$\text{{sgn}}$$\left(y-y'\right)/2$  (notice that we still refer to the antisymmetric in the band index bosonic field as "spin"). We rewrite the backscattering Hamiltonian (\ref{FBS}) in terms of the charge and spin bosonic fields as follows:
\begin{eqnarray}
H_{BS}&=&-\frac{2D}{\pi} \left(r_{s}\cos[\sqrt{2}\phi_{c}(0)]\cos[\sqrt{2}\phi_{s}(0)]\right.\nonumber \\
 &  & \left.+r_{a}\sin[\sqrt{2}\phi_{c}(0)]\sin[\sqrt{2}\phi_{s}(0)]\right.\nonumber \\
 &  & \left.+r_{ms}\cos[\sqrt{2}\phi_{c}(0)]\cos[\sqrt{2}\theta_{s}(0)]\right.\nonumber \\
 &  & \left.+r_{ma}\sin[\sqrt{2}\phi_{c}(0)]\sin[\sqrt{2}\theta_{s}(0)]\right),
\end{eqnarray}
where $r_{s}=||r_{+}|+|r_{-}||/2$, $r_{a}=||r_{+}|-|r_{-}||/2$,
$r_{ms}=||r_{\pm}|+|r_{\mp}||/2$, $r_{ma}=||r_{\pm}|-|r_{\mp}||/2$. 

\subsection*{Backscattering: Majorana fermions}

{\color{black} As a first step} we replace the charge mode by its mean value averaged over fast charge
degrees of freedom {\color{black} using the functional integral technique developed in Ref. \onlinecite{MA_theory}} {\color{black} and obtain the 
Hamiltonian:} 
\begin{eqnarray}
 &  & H_{\tau}(t)=\frac{v_{F}}{2\pi}\int_{-\infty}^{\infty}\{[\partial_{y}\theta_{s}(y)]^{2}+[\partial_{y}\phi_{s}(y)]^{2}\}dy\nonumber \\
 &  & -\sqrt{\frac{4D}{v_{F}}}\left(\omega_{s\tau}(t)\cos[\sqrt{2}\phi_{s}(0)]+\omega_{a\tau}(t)\sin[\sqrt{2}\phi_{s}(0)]\right.\nonumber \\
 &  & \left.+\omega_{ms\tau}(t)\cos[\sqrt{2}\theta_{s}(0)]+\omega_{ma\tau}(t)\sin[\sqrt{2}\theta_{s}(0)]\right),
\end{eqnarray}
where we use the notations (\ref{om},\ref{Omegas}) of the Section V.

{\color{black} As a next step we introduce the even and odd combinations of the "spin"  (aka sub-band)} bosonic fields
$\phi_{e/o}$$($$y$$)$$=$$[$$\phi_{s}(y)$$\pm$$\phi_{s}(-y)$$]$$/$$\sqrt{2}$,
$\theta_{e/o}$$(y$$)$$=$$[$$\theta_{s}(y)$$\pm$$\theta_{s}(-y)$$]$$/$$\sqrt{2}$. As a result, we obtain new chiral
fields $\Phi_{1/2}$$($$y$$)$$=$$\theta_{o/e}(y)$$-$$\phi_{e/o}$$($$y$$)$
satisfying the commutation relations: 
$[$$\Phi_{\alpha}(y)$$,$$\Phi_{\alpha'}(y')$$]$$=$$i$$\pi$$\delta_{\alpha\alpha'}$$\text{{sgn}}$$\left(y-y'\right)$ where $\alpha$$,$$\alpha'$$=$$1,2$.
We define new fermionic fields $\Psi_{\alpha}\left(y\right)$$=$$(\eta_{\alpha}/\sqrt{2\pi a})\exp{(-i\Phi_{\alpha}\left(y\right))}$
with a help of two local Majorana fermions $\eta_{1}$$=$$(d$$+$$d^{\dagger})$$/$$\sqrt{2}$
and $\eta_{2}$$=$$(d$$-$$d^{\dagger})$$/$$(i\sqrt{2})$ representing the quantum impurity \cite{MA_theory}.

Finally, we {\color{black} integrate out the fluctuations of the spin degree of freedom with  the frequencies exceeding $E_c$ (Ref. \onlinecite{MA_theory}). This procedure is equivalent to the poor man's scaling approach originally used for the Kondo problem \cite{Hewson} and leads to replacement of the bandwidth $D$ by the new bandwidth $T_K\sim E_c$.
As a result, we derive the effective Anderson model which} describes a hybridization of two local Majorana fermions $\eta_1$ and $\eta_2$ with two species of conduction electrons. The effective Hamiltonian (\ref{Kondo}) has a structure of two copies of the two-channel Kondo model where coupling constants $\omega_{i\tau}$ depend on {\color{black} both Zeeman and SOI fields}:
\begin{eqnarray}
 &  & H_{\tau}(t)=\int_{-\infty}^{\infty}dk
 \left[\sum_{\alpha=1,2} (k\cdot v_{F}) c_{\alpha,k}^{\dagger}c_{\alpha,k}\right.\label{Kon}\\
 &  & -\sqrt{2}\left.\left(\omega_{s\tau}(t)\eta_{1}(c_{1,k}-c_{1,k}^{\dagger})-i\omega_{a\tau}(t)\eta_{1}(c_{1,k}+c_{1,k}^{\dagger})\right.\right.\nonumber \\
 &  & \left.\left.+\omega_{ms\tau}(t)\eta_{2}(c_{2,k}-c_{2,k}^{\dagger})-i\omega_{ma\tau}(t)\eta_{2}(c_{2,k}+c_{2,k}^{\dagger})\right)\right].\nonumber 
\end{eqnarray}
\end{appendix}


\vspace*{-6mm}
\begin{thebibliography}{10}

\bibitem{LandauFL} L.D. Landau, Soviet Phys. JETP {\bf 3}, 920 (1957), ibid {\bf 5}, 101 (1957).

\bibitem{Anderson} P.W. Anderson, {\it The Theory of Superconductivity in the High-Tc Cuprate Superconductors}, Princeton University Press (1997).


\bibitem{HF} G. R. Stewart, Rev. Mod. Phys. {\bf 73}, 797 (2001).



\bibitem{Ralph} D. C. Ralph and R. A.  Buhrman, Phys. Rev. Lett. {\bf 69}, 2118, (1992);
D. C. Ralph, A. W. W.  Ludwig, J. von Delft and R. A. Buhrman, ibid {\bf 72}, 1064 (1994).


\bibitem{Goldhaber} R. M. Potok, I. G. Rau, H. Shtrikman, Y. Oreg
and D. Goldhaber-Gordon, Nature (London) \textbf{446}, 167 (2007).



\bibitem{NozBla80} P. Nozieres and A. Blandin, J.Phys (Paris) \textbf{41}, 193 (1980).

\bibitem{Zawad} A. Zawadowski, Phys. Rev. Lett. {\bf 45}, 211  (1980).

\bibitem{Affleck91} A.W.W. Ludwig and I. Affleck, Phys. Rev. Lett.
\textbf{67}, 3160 (1991).


\bibitem{Fazio} D. Boese and R. Fazio, Europhys. Lett.
\textbf{56}, 576 (2001).

\bibitem{Costi2010} T. A. Costi and V. Zlatic, Phys. Rev. B {\bf 81}, 235127 (2010).

\bibitem{Zitko} R. Zitko, J. Mravlje, A. Ramsak, T. Rejec, New J. Phys. {\bf 15}, 105023 (2013).


{\color{black}
\bibitem{Matveev95}
K.~A. Matveev, Phys. Rev. B {\bf 51},  1743  (1995).}

\bibitem{Flensberg} K. Flensberg, Phys. Rev. B {\bf 48}, 11 156 (1993).

\bibitem{FM_theory} A. Furusaki and K.~A. Matveev, Phys. Rev. Lett.
\textbf{75}, 709 (1995); Phys. Rev. B \textbf{52}, 16 676 (1995).

\bibitem{LeHur} K. Le~Hur, Phys. Rev. B \textbf{64},
161302(R) (2001), K. Le~Hur and G. Seelig, Phys. Rev. B \textbf{65},
165338 (2002).


\bibitem{MA_theory} A.~V. Andreev and K.~A. Matveev, Phys. Rev.
Lett. \textbf{86}, 280 (2001), K.~A. Matveev and A.~V. Andreev,
Phys. Rev. B \textbf{66}, 045301 (2002).


\bibitem{NKK10} T.K.T. Nguyen, M.N. Kiselev, and V.E. Kravtsov, Phys.
Rev. B {\bf 82}, 113306 (2010).


{\color{black} 
\bibitem{GG11} S. Amasha, I. G. Rau, M. Grobis, R. M. Potok, H. Shtrikman, D. Goldhaber-Gordon,  Phys. Rev. Lett. {\bf 107}, 216804 (2011). 
}



\bibitem{Mol1} R. Scheibner, H. Buhmann, D. Reuter,
M.~N. Kiselev, and L.~W. Molenkamp, Phys. Rev. Lett. \textbf{95},
176602 (2005).

\bibitem{Mol2} R.~Scheibner, E. G.~Novik, T. Borzenko,
M. Konig, D. Reuter, A.D. Wieck, H. Buhmann and L.~W. Molenkamp, Phys.
Rev. B \textbf{75}, 041301 (2007).


\bibitem{SOIbook} R. Winkler, {\it Spin-Orbit Coupling Effects
in Two-Dimensional Electron and Hole Systems}. (Springer, Berlin, 2003).

\bibitem{Spintronics} I. Zituc, J. Fabian, and S. Das Sarma, Rev.
Mod. Phys. \textbf{76}, 323 (2004).







\bibitem{GJPB05} V. Gritsev, G.I. Japaridze, M. Pletyukhov and D. Baeriswyl,
Phys. Rev. Lett. {\bf 94}, 137207 (2005).

\bibitem{Sun07} J. Sun, S. Gangadharaiah, and O.A. Starykh, Phys. Rev.
Lett. {\bf 98}, 126408 (2007); S. Gangadharaiah, J. Sun, and O.A. Starykh, Phys. 
Rev. \textbf{B 78}, 054436 (2008).

\bibitem{Miranda} R. G. Pereira and E. Miranda, Phys. Rev. B {\bf 71}, 085318 (2005).

\bibitem{Cheng} F. Cheng, K.S. Chan, and K. Chang, New J. of Phys,
\textbf{14}, 013016 (2012).



\bibitem{AG} I. L. Aleiner and L. I. Glazman, Phys. Rev. B \textbf{57},
9608 (1998).


{\color{black}
\bibitem{Kikoin12} K. Kikoin and Y. Avishai, Phys. Rev. {\bf B 86}, 155129 (2012).}


\bibitem{Aleiner01} I.L. Aleiner and V.I. Fal'ko, Phys. Rev. Lett.
{\bf 87}, 256801 (2001).




\bibitem{comSOI} We quantize the conduction electrons in the source using the SOI basis. The FL and tunneling Hamiltonians $H_s$ and $H_{tun}$ preserve the form under a replacement  $\sigma$$\to $$\nu$. 


\bibitem{Thierry}  T. Giamarchi, {\it Quantum physics in one dimension.}
(Clarendon; Oxford University Press, Oxford, 2004).

\bibitem{GNT} A. O. Gogolin, A. A. Nersesyan, and A. M. Tsvelik,
{\it Bosonization and Strongly Correlated Systems}. (Cambridge University Press, Cambridge, 2004).


{\color{black}


\bibitem{Cutler69} M.~Cutler and N.F.~Mott, Phys. Rev {\bf 181}, 1336
(1969).


\bibitem{Petta11} M. D. Schroer, K. D. Petersson, M. Jung, and J. R. Petta, Phys. Rev. Lett. {\bf 107}, 176811 (2011).

\bibitem{Mourik12} V. Mourik, K. Zuo, S.M. Frolov, S.R. Plissard, E.P.A.M. Bakkers, L.P. Kouwenhoven,
Science {\bf 336}, 1003 (2012). 

\bibitem{Hewson} A. C. Hewson, {\it The Kondo problem to Heavy Fermions}, (Cambridge University Press, Cambridge, 1993).

}
\end{thebibliography}
\end{document}